\documentclass[sigconf]{acmart}%,review
 
\AtBeginDocument{%
  }

\setcopyright{none}
\copyrightyear{2024}
\acmYear{2024}
\acmDOI{XXXXXXX.XXXXXXX}

\acmConference[SE 2030]{International Workshop on Software Engineering in 2030}{July 2024}{Porto de Galinhas, Brazil}
\newif\ifanonymize
\anonymizefalse  % Uncomment to show text

\begin{document}

\title{Contemporary Software Modernization: Perspectives and Challenges to Deal with Legacy Systems}

\ifanonymize

\author{Withheld for}
\affiliation{Double-blind
\country{review}}

\else

\settopmatter{authorsperrow=3}
\author{Wesley K. G. Assunção}
\orcid{0000-0002-7557-9091}
\affiliation{%
  \institution{North Carolina State University}
  \city{Raleigh} 
  \country{USA}
}
\author{Luciano Marchezan} 
\orcid{0000-0003-3096-580X}
\author{Alexander Egyed}
\orcid{0000-0003-3128-5427}
\affiliation{
\institution{Johannes Kepler University}
\city{Linz} 
\country{Austria}
}
\author{Rudolf Ramler}
\affiliation{
\institution{Software Competence Center Hagenberg GmbH (SCCH)}
\city{Hagenberg} 
\country{Austria}
}

\fi

\renewcommand{\shortauthors}{Assunção et al.}

\begin{abstract}
%Context: 
Software modernization is an inherent activity of software engineering, as technology advances and systems inevitably become outdated. The term ``software modernization'' emerged as a research topic in the early 2000s, with a differentiation from traditional software evolution. Studies on this topic became popular due to new programming paradigms, technologies, and architectural styles.
Given the pervasive nature of software today, modernizing legacy systems is paramount to provide users with competitive and innovative products and services.
%Problem: 
Despite the large amount of work available in the literature, there are significant limitations: (i) proposed approaches are strictly specific to one scenario or technology, lacking flexibility; (ii) most of the proposed approaches are not aligned with the current modern software development scenario; and (iii) due to a myriad of proposed modernization approaches, practitioners may be misguided on how to modernize legacies.
%Goal: 
In this work, our goal is to call attention to the need for advances in research and practices toward a well-defined software modernization domain. The focus is on enabling organizations to preserve the knowledge represented in legacy systems while taking advantages of disruptive and emerging technologies.
%Contributions: 
Based on this goal, we put the different perspectives of software modernization in the context of contemporary software development. We also present a research agenda with 10 challenges to motivate new studies.%, from some of which we present early results in our previous work.
\end{abstract}





\maketitle

% -- -- -- -- -- -- -- -- -- -- -- -- -- -- -- -- -- --
\section{Introduction}

Throughout the life of a software system, its architecture decays, its underlying technologies become obsolete, the user requirements change, or the company's business models evolve---ultimately, causing the software to morph into what we call legacy systems~\cite{Bennett1995}. 
The large majority of software currently in use are long-lived systems that represent many years of competitive knowledge and business value~\cite{khadka2014}. However, due to extensive maintenance and obsolete technology, legacy systems are costly to maintain, more exposed to cybersecurity risks, less effective in meeting their intended purpose, and push up costs of digital transformation~\cite{GAO2019,Morris2021,Beach2018}. 
For instance, the US government spent over \$90 billion in fiscal year 2019 on IT, from which about 80\% was used to operate and maintain legacy systems~\cite{GAO2019}.
Also, the UK government spends £4.7 billion a year on IT across all departments, and £2.3 billion goes on patching up systems, some of which date back 30 years or more~\cite{Morris2021}. 

To remain competitive, companies must modernize their legacy systems, preserving the hard-earned knowledge acquired through many years of system development~\cite{Seacord2003, khadka2014, Wolfart2021}.
According to Seacord et al.~\cite{Seacord2003} ``\textit{Software modernization attempts to evolve a legacy system, or elements of the system, when conventionally evolutionary practices, such maintenance and enhancement, can no longer achieve the desired system properties.}'' 
% Software Modernization is a form of software evolution, together with maintenance and replacement,  where an existing system is re-engineered\footnote{Software re-engineering is the examination and alteration of a system to reconstitute it in a new form and the subsequent implementation of a new form~\cite{Chikofsky1990}. Usually the alteration is transparent to the final user.}
% or migrated\footnote{Software migration is transferring a existing software system to a new environment to provide elaborated methods and techniques allowing already established software systems to benefit from the advantages of new technologies~\cite{Ionita2012}. Usually, the benefits of the new environment is perceived by the final user.}
% to a modern architecture or platform.
% {\color{blue} Software Modernization refers to the conversion, rewriting or porting of a legacy system to a modern computer programming language, protocols, or hardware platform. }
The process of modernizing a legacy system leads to benefits such as easing engineering activities, satisfying user needs, achieving new business goals, or reducing costs~\cite{Seacord2003}.
Furthermore, modernization is a mean to leverage the digital transformation~\cite{Beach2018}, as it enables the use of emerging/disruptive technologies such as artificial intelligence, high-performance computing, cloud computing, IoT, robotics, and big data~\cite{leon2021}.

%\textbf{Problem}: 
In the literature, we can find different modernization strategies~\cite{leon2021,strobl2020}. For example, restructuring systems using components, adoption of aspect-oriented development, re-engineering of system variants into software product lines, migration to microservices, and supporting for new hardware, e.g., multi-core/GPUs devices. Even the software development process has been modernized, e.g., agile methods~\cite{martin2003agile} and DevOps~\cite{Cherinka2022}.
Additionally, modernization has different driving forces and impacts related to \textit{organizational}, \textit{operational}, and \textit{technological} aspects~\cite{Wolfart2021}. 
For instance, the modernization can focus on independence for agile teams, optimize the deployment, ease the inclusion of innovation, facilitate scalability, or explore new market segments~\cite{Wolfart2021,strobl2020}.

Despite the existing studies on the topic of modernization, covering different strategies and  aspects, there are still significant limitations and gaps in the state of the art and the practice: 
(i) existing approaches are too specific and typically imply to individual technologies or specific modernization scenarios only, without flexibility. This limits their usefulness,  reusability, or adaptation for different technologies or scenarios; 
(ii) proposed approaches typically are not aligned with each other---often providing fragments of modernization and at times even becoming outdated, as there is no contemporaneous body of knowledge on the fundamentals of software modernization; 
and (iii) the existence of several different modernization strategies, on one hand, offers a wide range of potential solutions, however, on the other hand, such diversity of strategies may misguide practitioners, providers, and researchers when looking for solutions for specific situations.

The studies that try to organize the existing pieces of work on software modernization have several limitations. They only present an overview of the state of the art~\cite{leon2021}; are based on few case studies or a subset of existing literature~\cite{strobl2020,m2017assessment,iosif2015}; are outdated regarding current emerging/disruptive technologies~\cite{iosif2015,fanelli2016,khadka2015,khadka2014}; partially cover the modernization life cycle, and rarely take into account organizational, operational, and technological aspects~\cite{Wolfart2021,leon2021}. 
As pieces of work span across many years and focus on modernizing for different purposes, there is a need for discussing modernization in the context of the contemporaneous software development.

%\textbf{Goal}: 
In this work, our goal is to call attention to the need for advances in research and practices towards
software modernization in the light of contemporary software development. The focus is to preserve knowledge represented in legacy systems while employing disruptive and emerging technologies to the benefit of users, companies, and society.
%\textbf{Contributions}: 
Based on that, we contextualize the different  perspectives  of  software  modernization and introduce a research agenda with 10 challenges to be taken into account. %, from some of which we present early results in our previous work.
Our contribution is to motivate the discussion on software modernization, present open challenges, and alert companies about risks in adopting solutions based on popularity or hypes.

% -- -- -- -- -- -- -- -- -- -- -- -- -- -- -- -- -- --
\section{Background and Related Work}
\label{related_work}

Seacord et al.~\cite{Seacord2003} presented software modernization as a remedy to face the legacy system crisis in the early 2000s. They discussed how to keep or add business value through modern technologies, reducing operational costs, and dealing with technical aspects, e.g., allowing better reuse and easier maintenance~\cite{Seacord2003}. However, their discussion is not totally aligned with current technological and operational advances of contemporary software engineering. 

\begin{figure}[tbp]
\centerline{\includegraphics[width=1\linewidth]{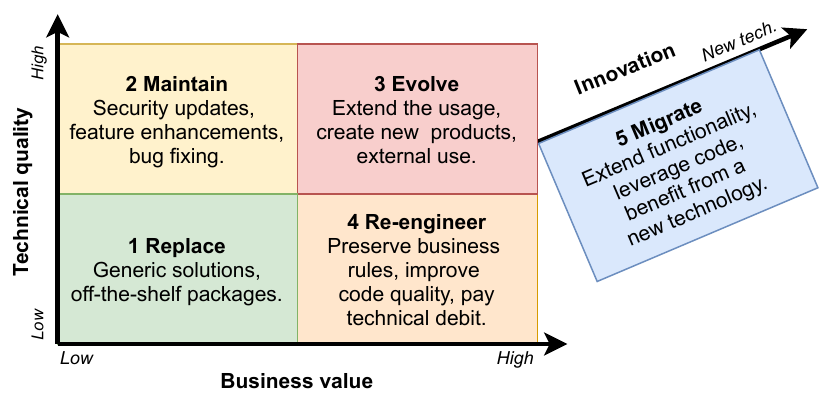}}
\caption{Extended quadrant of the portfolio analysis for the contemporary software modernization, adapted from~\cite{Seacord2001}.}
\label{fig:types_modernization}
\end{figure}

To decide for which modernization strategy to adopt, companies should perform a portfolio analysis. Figure~\ref{fig:types_modernization} presents the portfolio analysis quadrant extended from Seacord et al.~\cite{Seacord2003} to bring forward a contemporaneous perspective of software modernization. In addition to the \textit{technical quality} and \textit{business value} dimensions, we introduce \textit{innovation} as additional dimension that is achieved by new disruptive and emerging technologies---driving forces for the modernization.
The five quadrants presented in Figure~\ref{fig:types_modernization} are: 

\begin{itemize}
    \item \textbf{1 Replace}: legacy systems that have low business value and low technical quality, i.e., accumulated technical debt, should be replaced by new systems, using generic solutions or off-the-shelf systems, instead of undergoing a re-engineering or migration process.
    \item \textbf{2 Maintain}: systems with high technical quality and low business value should not require modernization effort, but traditional maintenance activities should be used, just to keep them operating and meeting customers need.
    \item \textbf{3 Evolve}: high-quality legacies with high business value should be actively evolved using traditional evolutionary development practices for introducing new features, new products, or even serving as third part for other systems.
    \item \textbf{4 Re-engineer}: systems with high business value and low technical quality should be re-engineered in order to preserve business value, i.e., external quality, and manage the technical debt, i.e., internal quality. This type of modernization can be transparent to the end user.
    \item \textbf{\textit{5 Migrate}}: when the system has high business value and a company decides to drive innovation with emerging or disruptive technologies, independently of the system's technical quality, a migration to the desired new technologies should take place. This is, for example, the case when companies foster a digitalization initiative.
\end{itemize}

In the literature, we can find \textit{several modernization strategies} to retain business value of legacy systems~\cite{leon2021, strobl2020}. For example, restructuring systems using components~\cite{Chiang2006, Ekanem2016, khalilipour2021}; adoption of aspect-oriented development~\cite{alsobeh2018aspect, rizvi2010comparative, goel2015legacy}; re-engineering of system variants into software product lines~\cite{Assuncao2017, Akesson2019, Kruger2020, Assuncao2020}; migration to microservices~\cite{Wolfart2021,Carvalho2019,Francesco2018industrial,Wang2021,taibi2017,Knoche2018,buchgeher2021adopting}; supporting new devices or pieces of hardware, e.g., from single-core to multi-core machines~\cite{Salman2021, Vinay2016, norton2008challenges}; classical information systems to quantum computing~\cite{Castillo2021,zhao2024unraveling}; and leveraging the use of AI/ML/Foundation Models~\cite{Agarwal2024}, a current trend. Even the software development process has been modernized, e.g., agile methods~\cite{martin2003agile} and DevOps~\cite{bommadevara2018,Cherinka2022}. %\lm{talvez aqui podemos introduzir a migracao de nao AI para AI como uma possibilidade. Esse paper por exemplo:}
Also, modernization has different driving forces and impacts related to \textit{organizational}, \textit{operational}, and \textit{technological} aspects. The modernization can focus on independence of teams, optimizing deployment, adding innovation, facilitating scalability, or exploring new market segment~\cite{Wolfart2021,strobl2020}.

Despite existing literature, the work on software modernization has several limitations: studies only present an overview of the state of the art~\cite{leon2021}; are based on few case studies or a subset of existing literature~\cite{strobl2020,m2017assessment,iosif2015}; are outdated regarding current emerging/disruptive technologies~\cite{iosif2015,fanelli2016,khadka2015,khadka2014}; partially cover the modernization life cycle, and rarely take into account \textit{organizational}, \textit{operational}, and \textit{technological} aspects~\cite{Wolfart2021,leon2021}. Furthermore, these studies are limited to exploring contemporary needs, e.g.,  digital transformation~\cite{leon2021}. Finally, software modernization must be seen as a  multi-perspective activity, which is discussed next.

% -- -- -- -- -- -- -- -- -- -- -- -- -- -- -- -- -- --
\section{Multi-perspective and Challenges}
\label{sec:challenges}

\begin{figure}[tbp]
\centerline{\includegraphics[width=.7\linewidth]{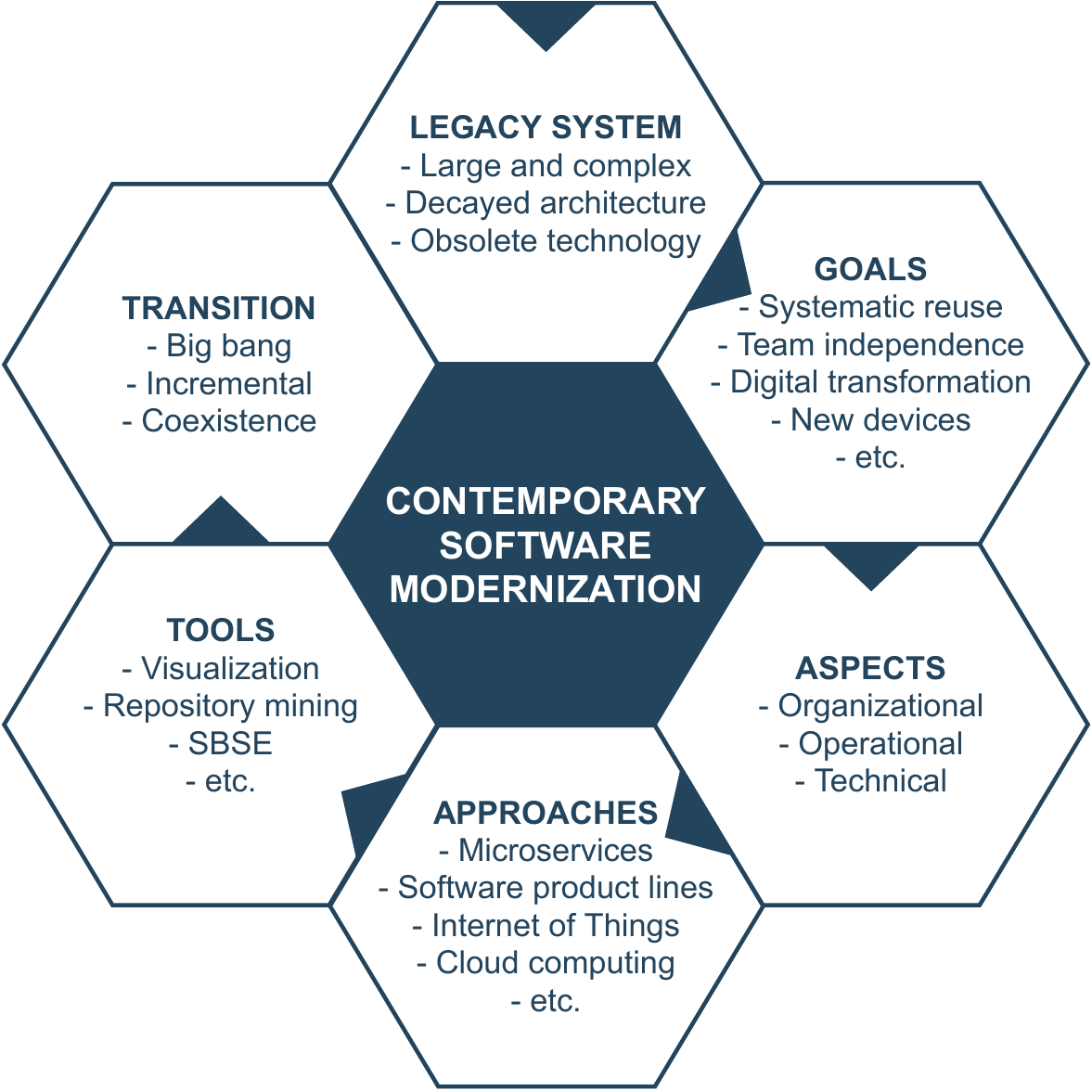}}
\caption{Different perspectives of software modernization in the context of contemporary software development.}
\label{fig:cont_mod}
\end{figure}

%Based on the state of the art and practice, and our previous work\lm{ref?}, we describe the
In this paper, we propose a multi-perspective of software  modernization in  the  context of contemporary software development. 
Figure~\ref{fig:cont_mod} presents six perspectives that affect the process of modernizing a legacy system. These perspectives range from understanding the legacy system, to conducting the transition from the legacy (or part of) to the modern system. %\lm{senti falta de uma discussao mais detalhada nas seis perspectivas} 
Based on this multi-perspective, together with known needs, trends, and recent pieces of work in the topic of software modernization, we present a research agenda structured as a list of challenges (C), described in what follows. %We have partially addressed some of these challenges, as we discussed in the next section. \textcolor{red}{Connect these challenges with Fig 1.}

\vspace{1mm} \noindent
\textbf{C$_1$: Lack of a comprehensive and contemporaneous body of knowledge on software modernization.}
The pieces of work that try to organize the existing body of knowledge on software modernization have several limitations, as discussed in Section~\ref{related_work}. Based on that, there is a need for a comprehensive and contemporaneous body of knowledge about  modernization strategies. We do not have to reinvent the wheel but organize existing knowledge in the light of the perspectives presented in Figure~\ref{fig:cont_mod} and contemporary software development approaches. Thus, we propose that researchers should gather and classify a body of knowledge on software modernization based on both research and practice. %\lm{Examples of initial effort?}
For example, a first step in that direction could be the knowledge base of architecture decision records collected from open source projects in \cite{buchgeher2023using}.

\vspace{1mm} \noindent
\textbf{C$_2$: Recommend the right approach based on the modernization goal.} In Figure~\ref{fig:cont_mod}, we present examples of goals that can serve as driving forces for modernization. Based on specific goals, some approaches are more appropriate than others. However, this recommendation must be an informed decision. The challenge here is to provide guidelines to support practitioners and companies on how to choose the proper approach according to their goals, avoiding deciding only based on technology ``hypes''. For example, microservice-based architectures have been advertised as a solution for technology flexibility. However, a recent study has shown that this is not the most common driving forces to migrate to microservices~\cite{Wolfart2021}. Choosing the wrong approach can lead to inefficiency and frustration in modernization. Hence, systems transformed into microservice were migrated back to monolithic applications~\cite{Mendonca2021}. Based on such experiences, future research should compile a body of knowledge (C$_1$) to derive guidelines for recommending and adopting customizable or domain-specific modernization approaches. %\lm{Examples of initial effort?} %\footnote{\url{https://changelog.com/posts/monoliths-are-the-future}}

% Avoid to only follow the “hype”. 
% Rafael Ponte, "Arquitetura Java: Escalando sua aplicação além do Hype" without microservices. 
% EASE paper that microservices are not for flexibility in technology. 
% Monolithic strikes back (monolithic is the future).

\vspace{1mm} \noindent
\textbf{C$_3$: Establish hybrid environments to allow the legacy and modern parts of a system operating together.} 
In Figure~\ref{fig:cont_mod}, we can see the three types of transition: (i) big bang, a.k.a. cold turkey~\cite{Comella2000}, which is the replacement of the legacy system with the modern one at once; (ii)~incremental modernization, following a strangler pattern~\cite{fritzsch2019}, in which parts of the legacy systems are incrementally replaced by modern parts~\cite{Seacord2003, Wolfart2021}; and (iii) the coexistence, in which legacy and modern parts operate together in one system~\cite{Robertson1997}.
The big bang and incremental transitions are explored in literature, however, little is discussed on how to establish a hybrid environment to allow the development and coexistence of legacy and modern system. In this context, we envision research to investigate how to enable hybrid environments, focusing on the coexistence of legacy and modern. This is related to C$_5$, as the decision made with regard to what to do with the legacy system.

%big bang\footnote{\url{https://www.cmswire.com/information-management/modernizing-legacy-tech-big-bang-or-piecemeal/}}

% \begin{figure}[!ht]
% 	\centering
% 	\includegraphics[width=.8\linewidth]{images/modernization_incremental.pdf}
% 	\caption{Incremental modernization of legacy systems, extracted from~\cite{Seacord2001}}
% 	\label{fig:mode_inc}
% \end{figure}

\vspace{1mm} \noindent
\textbf{C$_4$: Consider technical, operational, and organizational aspects during the modernization.} 
The great majority of studies on software modernization discusses the technical aspects of the modernization~\cite{Francesco2018industrial,Assuncao2017}. However, modernization has different driving forces and impacts related to \textit{organizational}, \textit{operational}, and \textit{technological} aspects~\cite{Wolfart2021,strobl2020, Seacord2001, Seacord2003}. %\lm{talvez citar exemploes dos impactos para cada categoria: org, ope, and tech} 
Software engineering involves technologies, people, and processes to be aligned with business strategies~\cite{Fitzgerald2017}. The challenge here is to propose approaches that deal with all these aspects. While technical aspects must always be considered, we argue that future work has to further explore organizational and operational aspects alongside the technical ones, balancing their priority and cost-benefit. %\lm{Examples of initial effort?}

\begin{figure*}[tbp]
\centerline{\includegraphics[width=.9\linewidth]{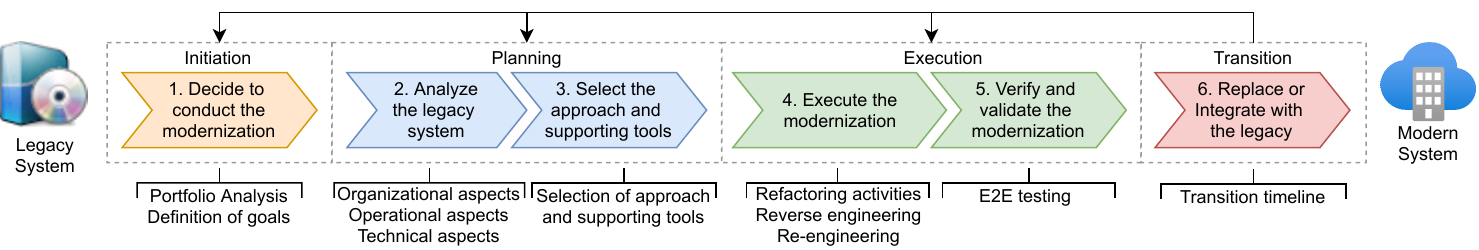}}
\caption{Preliminary multi-perspective modernization workflow in the context of contemporary software development.}
\label{fig:mod_process}
\end{figure*}

\vspace{1mm} \noindent
\textbf{C$_5$: Decide among replace, maintain, evolve, re-engineer, or migrate.} 
As presented in Figure~\ref{fig:types_modernization}, there are different forms of modernization. Also, in Figure~\ref{fig:cont_mod} we see that the legacy systems can present different problem related to its technical quality. Based on that, we can observe that choosing how to modernize a legacy system is a multi-criteria decision. Thus, companies need solutions to deal with this challenge. For that, we expect future work to propose recommendation approaches for decision-making support in terms of modernization possibilities, also taking into account organizational, operational, and technological aspects (C$_4$).

\vspace{1mm} \noindent
\textbf{C$_6$: Support digital transformation.} 
%According to Ebert and Duarte~\cite{ebert2018} ``\textit{Digital Transformation is about adopting disruptive technologies to increase productivity, value creation, and the social welfare.}'' 
Digital transformation is currently a trend and is receiving great attention around the world. 
For example, the European Union has the Digital Europe Programme,\footnote{\url{https://digital-strategy.ec.europa.eu/en/activities/digital-programme}}  %focused on bringing digital technology to businesses, citizens and public administrations, with a planned overall budget of EUR 7.5 billion.
Australia has the Digital Economy Strategy,\footnote{\url{https://digitaleconomy.pmc.gov.au/}}
%Australian government is investing almost AUD 1.2 billion in the Digital Economy Strategy\footnote{\url{https://digitaleconomy.pmc.gov.au/}} for building a modern and resilient economy.
in North America there are the Canada Digital Adoption Program\footnote{\url{https://www.ic.gc.ca/eic/site/152.nsf/eng/home}} and the Digital Strategy\footnote{\url{https://www.state.gov/digital-government-strategy/}} of the United States, and in Asia 11 countries have joined forces in the Connecting Capabilities.\footnote{\url{http://connectedfuture.economist.com/connecting-capabilities/}}
%The Canada Digital Adoption Program\footnote{\url{https://www.ic.gc.ca/eic/site/152.nsf/eng/home}} focus on support digitalization of small and medium-sized enterprises (SMEs).
%The United States have the Digital Strategy\footnote{\url{https://www.state.gov/digital-government-strategy/}} that aims to deliver better digital services to citizens.
%In Asia, 11 countries have joined forces in the Connecting Capabilities\footnote{\url{http://connectedfuture.economist.com/connecting-capabilities}}, a research programme exploring the potential for digital transformation. 
Despite expected benefits, the digital transformation is hampered by legacy systems~\cite{Beach2018}. In this context, modernization is a means to leverage the digital transformation~\cite{leon2021,Beach2018}. %, as it allows using emerging/disruptive technologies such as artificial intelligence, high-performance computing, cloud computing, IoT, robotics, and big data. 
However, there are no guidelines on how to perform software modernization to leverage digital transformation. Yet, the exiting few pieces of work on this topic only provide a superficial overview. In this direction, we envision future work enriching the legacy system with modern emerging and disruptive technologies to create new services and operations to companies' workforce and users.

\vspace{1mm} \noindent
\textbf{C$_7$: Prepare the legacy for the modernization.} 
When the legacy system has a high business value, it is a candidate for modernization by re-engineering or migration, independently of its internal quality (see Figure~\ref{fig:types_modernization}). However, understanding and modernizing a legacy system with poor internal quality is a complex task. For example, systems usually evolve in space, adding new features, and time, with features being revised~\cite{Michelon2021}, which make its comprehension difficult.
For such a situation, we believe that using refactoring strategies can be a good way to improve the legacy internal quality to face the modernization. However, the literature is scarce on how this ``pre-modernization'' activity should take place. This relates to the possibility regarding hybrid environments (C$_3$) as during the preparation for the evolution, the old legacy and the new migrated system may have to coexist. A research direction is to leverage Foundation Models in tasks related to code understanding~\cite{nam2024using} and refactoring~\cite{alomar2024refactor} during the modernization.

% Evolution in Space and Time?
% Refactoring? 
% Pre-modernization steps?

\vspace{1mm} \noindent
\textbf{C$_8$: Propose non-intrusive approaches and techniques.} 
Practitioners usually have preferences for using some technologies, tools, and workflows. %For instance, version control systems like Git are used by almost every developer. 
Based on that, researchers should propose modernization approaches and tools that take into account these preferences. Non-intrusive approaches and techniques are easier to transfer to practice~\cite{Costa2016}. Thus, we envisage that in addition to propose new solutions to the modernization challenges (e.g., C$_3$, C$_5$, C$_6$ and C$_7$), researches should also think of lightweight ways of integrating such solution in the technologies, tools, and workflows already in use by practitioners. This challenge is related to C$_4$, as the need for non-intrusive approaches reflects the need to consider operational and organizational aspects of companies.

% No intrusive tools?
% Visualization?
% \cite{Mortara2019, Liu2020}

\vspace{1mm} \noindent
\textbf{C$_9$: Train workforce with skills for dealing with modernization.} 
Figure~\ref{fig:cont_mod} presents the different perspectives of the software modernization. These several perspectives must be considered to train the workforce in charge of operationalizing the modernization process~\cite{prokofyev2019}. Thus, a challenge is to training the workforce with expertise to deal with the complexity of software modernization~\cite{Castillo2021}. To address this challenges, educators can benefit from contributions to C$_1$, in which a body of knowledge can serve as the basis for designing new courses in academia. Training also relates to C$_4$, as it directly affects operational and organizational aspects that are important for companies, such as empowering employees.

%\footnote{\url{https://www.forbes.com/sites/forbestechcouncil/2019/06/19/building-the-quantum-workforce-of-the-future/?sh=11aebf48fa47}}

% \vspace{1mm} \noindent
% \textbf{CX: Modernization of software from cross-domain systems.} 
% Automotive, robotics, cyberphysical systems.
% Collaborative?
% Cloud computing
% (Micro-)service architectures
% ML/AI-based systems
% Cyber-physical systems
% Robotics
% Internet of things
% Automotive software 
% Autonomous driving 
% Bots in software engineering

% \vspace{1mm} \noindent
% \textbf{CX: Variability management in modern architectures.} 
% SPLC challenge, workshop, Re-engineering legacy applications into SPLs~\cite{Assuncao2017}
% International Workshop on Variability Management for Modern Technologies (VM4ModernTech)
% https://sites.google.com/view/vm4moderntech-2021

\vspace{1mm} \noindent
\textbf{C$_{10}$: Modernization for small and medium-sized enterprises (SMEs).} 
In the literature, we observe that some software engineering activities should be conducted differently in the context of SMEs~\cite{Sanchez2016, Silva2014}. This might also be the case for software modernization~\cite{Althani2016}. Based on that, research needs to be conducted to deal with challenges faced by SMEs when modernizing their legacy systems to grow and be more competitive. This should be considered by researchers when creating a body of knowledge for software modernization. The modernization in context of SMEs also relate directly to C$_4$, as the organizational aspects of SMEs are different from big enterprises.  Consequently, it may have a significant impact on the decision to modernize the legacy or not (C$_5$).

% -- -- -- -- -- -- -- -- -- -- -- -- -- -- -- -- -- --
\section{The Modernization Workflow}

Based on recent systematic mapping studies~\cite{Wolfart2021,Assuncao2017}, we defined a preliminary multi-perspective modernization workflow, which is presented in Figure~\ref{fig:mod_process}. This process is composed of four phases, namely initiation, planning, execution, and transition. Additionally, these phases have six activities. The activities are sequential, but it is possible to return to previous activities, as shown by the arrows. Below each activity, we describe some tasks, which are related to the modernization quadrant (Figure~\ref{fig:types_modernization}) and the multi perspectives (Figure~\ref{fig:cont_mod}). 
This workflow is an initial proposal for establishing a general process to be enriched with specific information or additional activities to deal with the challenges presented in Section~\ref{sec:challenges}. For instance, if a process is desired to SMEs (C$_{10}$), constraints related to limited resources should be considered. 
%If the goal is modernizing a set of legacy system variants to systematically deal with variability, an activity for the analysis of similarities and commonalities among variants can be included.
%We proposed a modernization approach to migrate software variants to a systematic-reuse platform using UML models to avoid being dependent of a programming specific language~\cite{Assuncao2020IST} (CX). Variability debt, Evolution Space and time of HCSSs~\cite{MichelonGPCE2021, MichelonSPLC2020} Combination of static and dynamic analysis to support re-engineering of legacy systems to SPLs

% Generic strategy to migrate legacy applications to microservices considering multiple operational and technical aspects~\cite{Assuncao2021SANER,Carvalho2020ICSME,Assuncao2021EMSE}. 
% Based on a systematic mapping study, we proposed a modernization process for modernizing legacy systems with microservices~\cite{Wolfart2021}. This process was evaluated by practitioners to investigate if research and practice are aligned~\cite{Colanzi2021SBCARS}

% Tools for refactoring~\cite{Tenorio2020SBES, Bibiano2021ICSME, Uchoa2021MSR}

% Visualization~\cite{Crescencio2019JBCS}

% -- -- -- -- -- -- -- -- -- -- -- -- -- -- -- -- -- --
\section{Conclusion}

Software modernization is a fundamental activity of software engineering, since inevitably requirements change, and technology advances, and new business models emerge. Despite that, research on this topic has not been following the modern software development, and legacy systems still remain a problem. 
To fill this gap and to sparkle the research on this topic, we present a discussion of software modernization in the light of contemporary software modernization. We revisited some pieces of work and introduce the multi-perspective of contemporary software modernization. 
Based on that, in this work, we discussed 10 challenges to motivate and guide to new studies. These challenges can be employed as a research agenda for future work on this topic.
%a systematic mapping and our experience on software modernization, 
% We discussed 10 challenges to motivate and guide to new studies. 
% Finally, as an early achievement of our ongoing work, we distilled a preliminary multi-perspective modernization process to serve as basis for instantiations to meet modernization needs.


\ifanonymize
\begin{acks}
Withheld for double-blind review.
\end{acks}
\else
\begin{acks}
The research reported in this paper has been funded by BMK, BMDW, and the State of Upper Austria in the frame of SCCH, part of the COMET Programme managed by FFG as well as the Austrian Science Fund (FWF, P31989-N31).
\end{acks}
\fi

\bibliographystyle{ACM-Reference-Format}
\balance

\end{document}